\documentclass[printer]{aa}
\usepackage{graphics,graphicx}  

\begin{document}

\title{Lopsided spiral galaxies:\\evidence for gas accretion}
                                
\author{F. Bournaud \inst{1}
\and F. Combes \inst{1} 
\and C. J. Jog \inst{2}
\and I. Puerari \inst{3}
} 
\offprints{F. Bournaud, \email{frederic.bournaud@obspm.fr}}

\institute{Observatoire de Paris, LERMA, 61 Av. de l'Observatoire, 
   F-75014, Paris, France 
\and Department of Physics, Indian Institute of Science, 
  Bangalore 560012, India
\and Instituto Nacional de Astrof\'\i sica, Optica y
   Electr\'onica, Calle Luis Enrique Erro 1, 
   72840 Tonantzintla, Puebla, M\'exico
}
 
\date{Received XX XX, 2004; accepted XX XX, 2004}
\authorrunning{Bournaud et al.} 

\abstract{We quantify the degree of lopsidedness for a sample of 149 galaxies observed in the near-infrared from the OSUBGS sample, and try
to explain the physical origin of the observed disk lopsidedness. 
We confirm previous studies, but for a larger sample, that
a large fraction of galaxies have significant lopsidedness in their stellar disks, measured as the Fourier amplitude of the $m=1$ component normalised to the average or $m=0$ component in the surface density.
Late-type galaxies are found to be more lopsided, while the presence of $m=2$ spiral arms and bars is correlated with disk lopsidedness. We also show that the $m=1$ amplitude is uncorrelated with the presence of companions. Numerical simulations were carried out to study the generation of $m=1$ via different processes: galaxy tidal encounters, galaxy mergers, and 
external gas accretion with subsequent star formation. These simulations show that galaxy interactions and mergers can trigger strong lopsidedness, but do not explain several independent statistical properties of observed galaxies. To explain all the observational results, it is required that a large fraction of lopsidedness results from cosmological accretion of gas on galactic disks, which can create strongly lopsided disks when this accretion is asymmetrical enough.

}

\maketitle

\section{Introduction \label{intro}} 

It has been known for a long time that the gaseous component (HI)
in galaxies is often strongly asymmetric and lopsided.
Baldwin et al. (1980) have first studied this remarkable phenomenon in detail and proposed that the perturbations are $m=1$ waves 
built from off-centered elliptical orbits that may persist a long time
against differential precession. However, they have derived a
lifetime that is still not  sufficient to explain the high frequency of lopsidedness in neutral gas disks.
Richter \& Sancisi (1994) have compiled a sample of 1700 galaxies
from the literature and noticed on resolved maps of nearby galaxies (like M101)
that there is a correspondence between lopsidedness and
the global HI velocity profile of the galaxy. It is therefore possible to
study the asymmetries directly on the global HI spectrum, for a much larger sample. They derive 
 a lower limit of 50\% for the fraction of lopsided galaxies, and conclude that
HI asymmetries in disk galaxies are the rule rather than the exception.
Lopsidedness is more frequent in late-type galaxies, where
Matthews et al. (1998) found a frequency of 77\% of HI-distorted
profiles. However, it is difficult to get a precise quantitative indicator of the degree of lopsidedness with HI profiles without any spatial resolution. 

The ubiquity of lopsidedness may not be the privilege of HI disks. The stellar
disk potential, traced by near-infrared light, can be strongly 
lopsided, even in isolated galaxies, like the spectacular case of NGC 1637 (Block et al. 1994). The stellar light is definitely affected by
the phenomenon, as shown by the studies of 
Rix \& Zaritsky (1995) and Zaritsky \& Rix (1997) with a total
sample of 60 galaxies. They computed the $m=1$ Fourier amplitude of the density, and found that at least one third of these galaxies are 
significantly lopsided. Kornreich et al. (1998) quantify the
asymmetry by comparing the optical light contained within trapezoidal sectors
of galactic disks, and find also that 30\% of galaxies are very lopsided. In addition to the spatial asymmetry discussed above, some galaxies also display kinematical lopsidedness though this has been less studied (Schoenmakers et al. 1997, Swaters et al. 1999).

To explain the high frequency of lopsidedness, two kinds of arguments have been
proposed. First the perturbation might be longer-lived than previously thought,
either because the winding out by differential precession is quite long
in the outer parts of galaxies (Baldwin et al. 1980), because of weakly damped modes (Weinberg 1994), because of the amplification of density waves (Shu et al. 1990, Syer \& Tremaine 1996), or alternatively that the disk is distorted in a lopsided halo (Jog 1997, 2002) or else off-centered with respect to a massive halo (Levine \& Sparke 1998, Noordermeer et al. 2001). Second, the perturbations could be forced by tidal interactions (Kornreich et al. 2002) or minor mergers (Zaritsky \& Rix 1997).
In the last cases, the perturbations could be short-lived, and either lopsided
galaxies should be observed with companions or the merger frequency is very high.
For instance, Walker et al. (1996) have shown through N-body simulations
that a minor merger can result in a lopsided disk with perturbations lasting at least 1 Gyr. 
But until now, there has been no observed strong correlation
between the presence of companions and lopsidedness
(Wilcots \& Prescott 2004). Thus, what generates the large fraction of lopsidedness in the disks of spiral galaxies remains unknown.

In this paper, we measure the lopsidedness for a larger statistical sample of galaxies, and try to understand the
physical origin of the observed disk lopsidedness.
In Sect.~\ref{obs} we quantify the $m=1$ Fourier amplitude of the surface density in a sample of 149 galaxies with available near-infrared images from the Ohio State University Bright Galaxy Survey (OSUBGS, Eskridge et al. 2002). Since the galaxies chosen are not lopsided a priori, this sample is expected to
give a definitive measure of lopsidedness in typical disk galaxies. We also measure the HI asymmetry for the present sample of galaxies. The dependence of the degree of lopsidedness versus other physical parameters (presence of companions, bars and spiral arms, and Hubble type) is studied in Sect.~3. In order to understand the origin of the disk lopsidedness, we examine various scenarios via numerical simulations: distant tidal encounters, minor mergers, and gas accretion (Sect.~4). We show that galaxy interactions and mergers are not enough to explain all the properties of the observed galaxies. Cosmological accretion of gas seems to be responsible for a large fraction of lopsided disks. Observations of lopsided disks imply that galaxies have accreted large amounts of gas in past Gyrs, as already expected from the study of barred galaxies (Bournaud \& Combes 2002, Block et al. 2002). These results are discussed and we give our conclusions in Sect.~5.

\section{Analysis of observations \label{obs}}

\subsection{The OSUBGS sample}

To quantify the amplitude of the $m=1$ mode both in the disk surface density and in the potential, we use near-infrared images from the
Ohio State University Bright Galaxy Survey (OSUBGS, Eskridge et al.
2002).  This is a magnitude-limited sample drawn from the {\it Third
Reference Catalog of Bright Galaxies} (RC3, see Vaucouleurs et al. 1991).
For a spiral galaxy to be included in OSUBGS sample, its $B$
magnitude has to be brighter than 12$^{m}$, and its diameter less
than $6\arcmin$. According to the RC3, the catalog is at least 95\% complete
for galaxies this bright. We use the same techniques for deriving 
the various $m$-Fourier amplitudes as in Block et al.
(2002).

From the 174 galaxies in the sample, 171 are appropriate 
for studies of surface density distribution. We include only galaxies 
that are inclined up to 70$^\mathrm{o}$ with respect to the line of sight, which gives a total of 149 galaxies. The choice of this inclination cut-off is 
justified in Sect.~2.3.

\subsection{The bulge/disk decomposition and deprojection}

The bulge and disk components are decoupled through radial profile
Fitting; the bulge in the center is fitted by an $r^{1/4}$ de Vaucouleurs
profile, and the disk by an exponential profile. 

To deproject the galaxies, a particular 3D shape has to be assumed
for each component. We assume the disk is flat, with constant
scale-height (which will be varied next). Bulge components are likely to be flattened, but less than disks, and the flattening of the bulge is quite difficult to derive for non edge-on galaxies. In a first study, we do not subtract the bulge in the image, and deproject it together with the disk, assuming that the bulge is as flattened as the disk. In a second study we subtract the bulge light from the image, before deprojecting the disk, and then we add it to the deprojected image, assuming the bulge is spherical. We believe that the two solutions will bracket the actual one, so we record the average value. However we checked that choosing one of the two solutions does not result in any significant change in our analysis of lopsidedness: in the following we mainly study the disk surface density outside of the bulge radius, so that assumptions about the bulge component are not critical. 

A constant mass-to-light ratio and a constant scale-height are assumed. The
 value of $m=1$ amplitude in surface density is computed for three values 
 of the scale-height (225, 325, and 425 pc) and we keep the average value.

\subsection{The $A_1$ parameter: Lopsidedness in surface density}\label{a1def}

\subsubsection{Definition}

The near-IR light is assumed to be the best tracer of old stars that dominate the stellar mass, less affected by dust extinction or star formation. 
We compute the Fourier transforms of the stellar surface density, taken
from the NIR images.

The surface density is decomposed as:

\begin{equation}
\mu(r,\phi) = \mu_0(r) + \sum_m a_m(r) \cos (m \phi - \phi_m (r))
\end{equation}
where  the normalised strength of the Fourier component $m$ is 
$A_m(r) =  a_m / \mu_0 (r)$. Thus, $A_1$ represents the normalized 
amplitude of the disk lopsidedness at a given radius, and $A_2$
represents the same for an $m=2$ distortion like a bar or spiral arms. The quantity $\phi_m$ denotes the position angle or
the phase of the Fourier component $m$. 
We further define $<A_1>$ as the average of the normalized amplitudes
$A_1 (r)$ between 1.5 and 2.5 exponential radii.

Figure~1 shows a plot of the amplitude $A_1 (r) $ and the phase
$\phi_1 (r)$ versus radius $r$ for NGC~1637, a typical lopsided galaxy (Block et al. 1994); the deprojected density map computed from the near-IR image from the OSUBGS data is shown later in Fig.~\ref{nac}. 
This illustrates the main features of disk lopsidedness, namely that
$A_1$ increases with radius, and the phase is fairly constant
with radius in the outer parts. This is the case for most galaxies in our sample, even though there are a few exceptions, and it confirms the previous result by Rix \& Zaritsky (1995) for a larger sample studied here.

\begin{figure}
\rotatebox{-0}{\includegraphics[width=8cm]{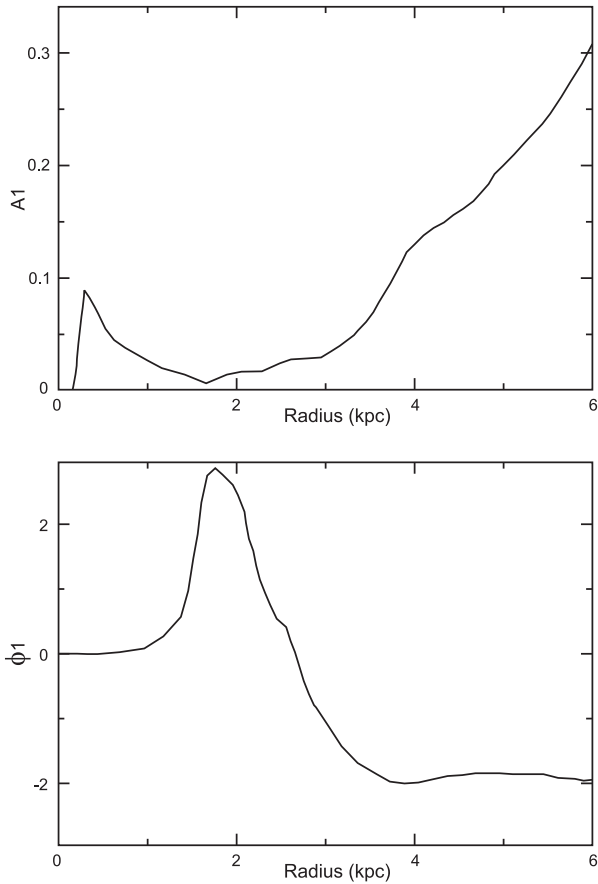}}
\caption{Radial variation of $A_1$, the normalized Fourier amplitude of the
stellar density, and phase $\phi_1$, for the
lopsided galaxy NGC 1637, measured from the near-infrared image
of the OSUBGS sample. The disk scale-length is 2.45 kpc and the value of $<A_1>=0.19$ was measured between 3.7 kpc and 6.1 kpc. The near-IR image of the
galaxy from the OSUBGS catalog is shown in Fig.~\ref{nac}. This galaxy shows a high degree of lopsidedness, but has no interacting companion and no evidence of a recent merger.}
\label{ngc1637}
\end{figure}

The distribution of $A_1$ values vs. radius for the 163 OSU
sample of galaxies (with inclination $< 70^\mathrm{o}$) is given in the Appendix. The average values $<A_1>$ and $<A_2>$ for the sample are listed in Table~\ref{TabA} of Appendix~\ref{app}.

\subsubsection{Statistical properties}

The histogram of the $<A_1>$ parameter is displayed in 
Fig.~2, and the cumulative function\footnote{The cumulative function of $<A_1>$ displays the fraction of galaxies with $<A_1>$ larger than the x-axis value} of $<A_1>$ in Fig.~3. The histogram confirms and extends the previous results by Rix \& Zaritsky (1995) for a larger sample. A good threshold for deciding that a galaxy is ''lopsided'' is $<A_1>=0.05$: according to our numerical simulations in Sect.~4, a value of $<A_1>$ larger than 0.05 is unlikely to result from internal mechanisms. Internal mechanisms may in theory trigger $m=1$ asymmetries (e.g., Masset \& Tagger 1997), but our simulations shown in Sect.~4, as well as other existing works, do not show strong lopsidedness arising spontaneously without any external perturbation. With this criterion, we find that 63\% of the galaxies are lopsided. The mean value of $<A_1>$ over our sample is 0.11, and 34\% of galaxies have $<A_1>$ larger than this average value.

\begin{figure}
\rotatebox{-0}{\includegraphics[width=8cm]{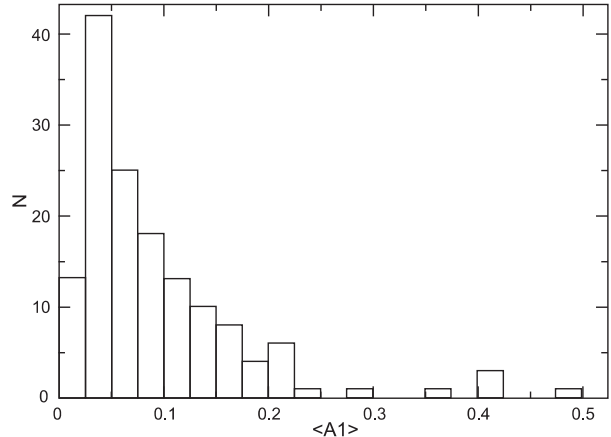}}
\caption{Histogram of $m=1$ Fourier amplitude in the
stellar density deduced from the near-infrared images of the OSUBGS sample of galaxies. The $<A_1>$ parameter is the average of the normalized
Fourier amplitude for $m=1$, over the radial range between 
1.5 to 2.5 disk scalelengths. 
\label{histogram}}
\end{figure}

In the above analysis, we have restricted the sample to galaxies inclined by less than $< 70^\mathrm{o}$. This value is
chosen because the statistical distribution of the $<A_1>$ parameter 
is nearly identical whether the cut-off is chosen to be $30^\mathrm{o}, 50^\mathrm{o}$, or $70^\mathrm{o}$ (see Fig.~3), and the higher value in this range was chosen as it gives a larger sample which is more useful for a statistical study. For higher values of the cut-off, statistical properties appeared to be changed, for the lopsidedness can no longer be correctly estimated.

\begin{figure}
\rotatebox{-0}{\includegraphics[width=8cm]{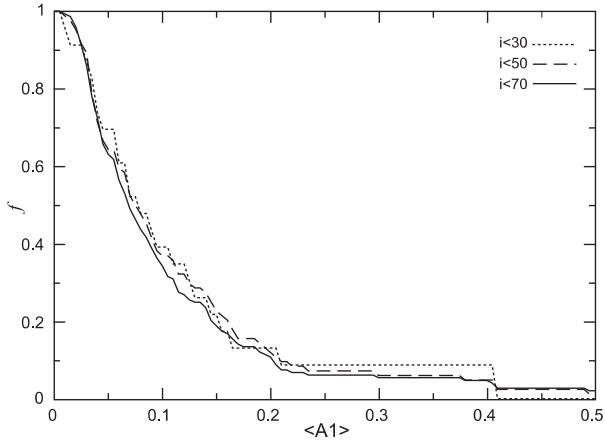}}
\caption{Cumulative functions of $<A_1>$ (fraction of galaxies with $<A_1>$ larger than the x-axis value) for galaxies seen below a certain cut-off in
the inclination angle $i$; for $i < 30^\mathrm{o}, 50^\mathrm{o},$ and $70^\mathrm{o}$.
The curves are nearly identical, hence the highest value of the
cut-off $70^\mathrm{o}$ is chosen, as it gives the largest number of
galaxies in the sample. Galaxies more inclined than $70^\mathrm{o}$ no longer follow the same statistical distribution and have been rejected.
\label{cutoffnb}}
\end{figure}

\subsection{The $Q_1$ parameter}

We next derive the gravitational potential of each galaxy from the near-infrared images. This potential gives a global
estimation of the perturbation of the galaxy disks, but is weighted
quite differently from the density distribution. This gives 
a complementary view of the perturbations, weighted more
globally, while the $A_1$ parameter gives values which are weighted
more locally.

The image of the galaxy is completed in the 3rd dimension, assuming
an isothermal plane with a scale-height constant with radius and equal to 1/12th of the radial scale-length. The gravitational potential is deduced by the Fourier transform method on the grid assuming a constant mass-to-light ratio. 

The potential is decomposed as:
\begin{equation}
\Phi(r,\theta) = \Phi_0(r) + \sum_m \Phi_m(r) \cos (m \theta - \phi_m)
\end{equation}
We define the strength of the $m$-Fourier component,
$Q_m(r) = m \Phi_m / r | F_0(r) |$, and its global strength
over the disk as $\max_r \left( Q_m(r) \right)$ (e.g., Combes \& Sanders 1981). 

The maximal torque at a given radius is defined by
\begin{equation}
Q_T(R) = {F_T^{max}(R) \over F_0(R)} = 
{{{1\over R}\bigl{(}{\partial \Phi(R,\theta)\over \partial\theta}\bigr{)}_
{max}} \over {d\Phi_0(R)\over dR}} 
\end{equation}
\noindent
where $F_T^{max}(R)$ represents the maximum amplitude of the tangential force at radius $R$, and $F_{0}(R)$ is the mean axisymmetric radial force inferred from the $m$=0 component of the gravitational potential. Our interest is focussed here on the $m=1$ component, $Q_1$, defined as the maximum of $Q_1(R)$ between 1.5 and 2.5 disk scale-lengths.

We display the histogram of the values of $Q_1$ in Fig.~4. This is similar to the histogram for $<A_1>$ as expected. We prefer to use $A_1$, the Fourier amplitude of lopsidedness in surface density, as the main indicator of disk lopsidedness since it is a directly measured quantity, and is frequently used in the literature.

\begin{figure}
\rotatebox{-0}{\includegraphics[width=8cm]{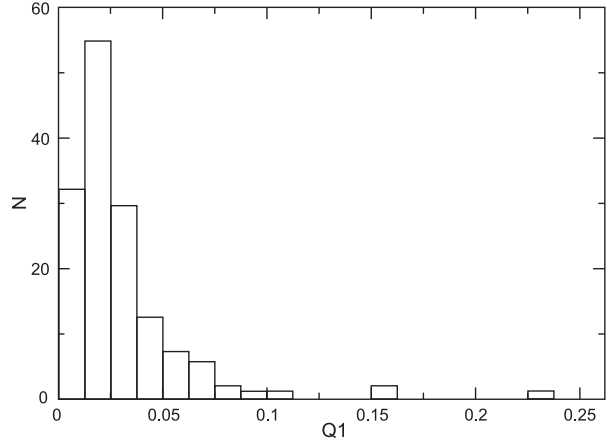}}
\caption{Histogram of $m=1$ Fourier amplitude in the
gravitational potentials ($Q_1$) for the OSUBGS sample of galaxies. 
\label{Q1max}}
\end{figure}

\subsection{Lopsidedness of gaseous disks\label{HI}} 

The asymmetries in the HI component have been estimated in a large sample on global velocity profiles by Richter \& Sancisi (1994). Since galaxies with interferometric HI maps are rare, they investigate the usability of global HI profiles as tracers of the lopsidedness of the HI disks in galaxies and notice that asymmetries in the global profiles usually correlate with those found on the HI maps. There are, however, cases where the HI map asymmetry does not translate into the HI global profile asymmetry, because of projection effects of the velocity along the line-of-sight, or because the asymmetry changes sign over the disk and conspires to reduce the global projected the perturbation. Therefore the velocity profile tracer is likely to underestimate the asymmetry level.

\begin{figure}
\centering
\resizebox{8cm}{!}{\includegraphics{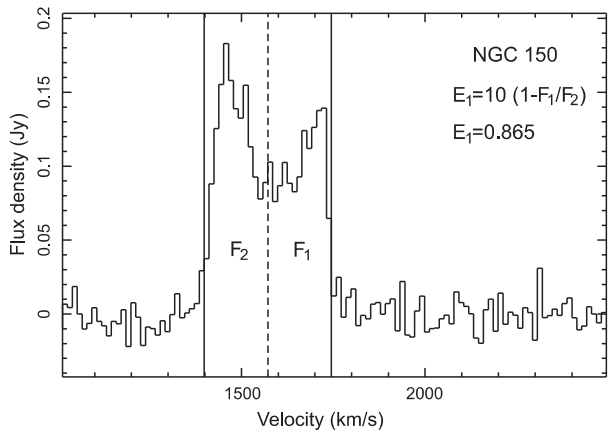}}
\caption{HI velocity profile of NGC~150, showing the two horns, the center of the profile and the extremity of the two horns. F$_1$ and F$_2$ are the integrated flux of each horn, and their ratio gives an estimate of the HI asymmetry.}\label{HIprof}
\end{figure}

We gathered HI velocity profiles from the literature
for our present sample, in order to estimate the neutral gas
lopsidedness. The HI asymmetry was measured in terms of the second criterion from
Richter \& Sancisi (1994). We first compute the flux ratio $f$ 
between the two halves of the profiles (the two horns), as illustrated in 
Fig.~\ref{HIprof}, to compare the mass of the approaching side of 
the galaxy, and the mass of the receding side. To measure the flux of each 
horn, we took the total width at 20\% of peak flux density, and determined both 
the extremities of the profile and its center, i.e. the division between the 
two halves.

\begin{figure}
\centering
\resizebox{8cm}{!}{\includegraphics{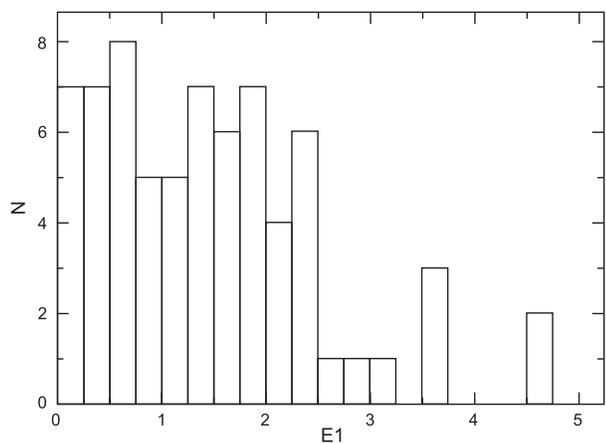}}
\caption{Histogram of the $E_1$ parameter measured on HI velocity profiles.}\label{HIhisto}
\end{figure}

We define the asymmetry parameter as $E_1=10(1-f)$, where 
$f$ is the flux ratio of the two horns of the profile. 
HI data available for 76 galaxies of our sample. The corresponding histogram 
of $E_1$ is shown in Fig.~\ref{HIhisto}, which shows that asymmetries in the HI component are very frequent. The asymmetry is larger than 10\% ($E_1$ larger than 1) for 
nearly two thirds of the galaxies in our sample, while in the stellar component 
only one third of galaxies have $<A_1>$ larger than 0.1. Asymmetries of 
more than 20\% ($<A_1>>0.2$ or $E_1>2$) are also much more common in the 
HI component than in NIR disks. Even if $E_1$ only provides a lower limit to the actual HI asymmetry, we find that in our sample lopsidedness is even more frequent and stronger in gas than in stars.

\section{Dependence of the lopsidedness on various physical parameters}

In this section, we study the statistical relations between lopsidedness and other parameters (the Hubble type, the presence of companions, and the presence of $m=2$ asymmetries like arms or bars). The results presented here are for the sample of galaxies below the inclination cut-off of $70^\mathrm{o}$, but we made sure that the results are not artifacts of deprojection by checking that they remain unchanged for a cut-off of $50^\mathrm{o}$ and even for a cut-off of $30^\mathrm{o}$ when the number of galaxies below $30^\mathrm{o}$ was large enough to make this study possible.

\subsection{Dependence of $<A_1>$ on galaxy type}

To study the dependence of the lopsidedness $<A_1>$ on
galaxy types, we plot the cumulative functions of $<A_1>$ for three groups of spiral galaxies in Fig.~7 : the early-types ($0 < T < 2.5$), the intermediate-types ($2.5 < T < 5$), and the late-types ($5 < T < 7.5$). The correspondence between the Hubble type parameter $T$ and the Sa--Sd classification of early-type/late-type spiral galaxies, as explained for instance by Binney \& Merrifield (1998), is: $T\simeq 2$ for Sa, $T\simeq 4$ for Sb, $T\simeq 7$ for Sc. The number of galaxies with $T>7.5$ in our sample is not large enough to give any representative result on the 
extremely late-type systems.

\begin{figure}
\rotatebox{-0}{\includegraphics[width=8cm]{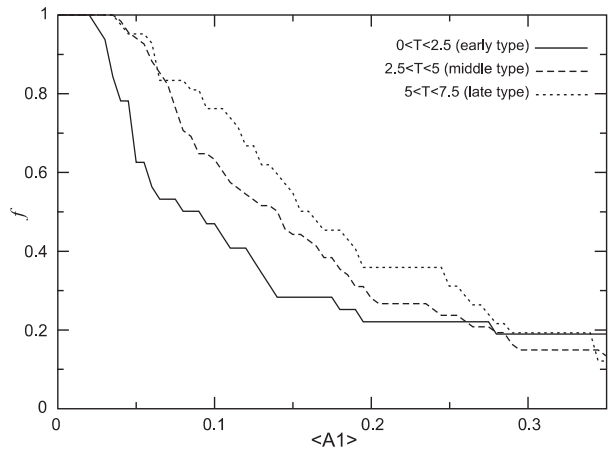}}
\caption{Cumulative function of $<A_1>$, for three groups of Hubble types of spiral galaxies: the
early-types ($0 < T < 2.5$), the intermediate-types 
($2.5 < T < 5$), and the late-types ($5 < T < 7.5$).
The late-type spirals are more lopsided than the early-type ones. 
\label{hubbletype3}}
\end{figure}

The late-type galaxies have a higher fraction of lopsided galaxies, as can be seen for any value of lopsidedness $<A_1>$. This result is similar to what Matthews et al. (1998) found in HI data. 

If we had employed $Q_1$ instead of $A_1$, we could think that this correlation between the asymmetry and the Hubble type is an artefact of bulge contamination, since the axisymmetrical contribution of the bulge to the potential reduces the value of $Q_1$ in early-type spiral galaxies more than in late-type galaxies, even beyond the bulge radius. However, since we have used the $A_1$ parameter and taken its average value between 1.5 and 2.5 disk scale-length, our results are not affected by bulge contamination, except maybe in a few cases were the bulge has a large radial extent, but this is rare. Then, the correlation between $<A_1>$ and the Hubble type really means that late-type spirals are more lopsided than early-type ones.

\subsection{Comparison of $<A_1>$ and $<A_2>$}

In Fig.~8, we show the cumulative distribution of $<A_1>$ for galaxies that have strong and weak $m=2$ asymmetries, and find that there is a clear correlation between $A_1$ and $A_2$. A similar result is obtained when we measure $<A_2>$ as the average value between 0.5 and 1.5 disk scalelengths or between 2.5 and 3.5 disk scalelengths, suggesting that the presence/strength of a lopsidedness is correlated to the presence/strength of $m=2$ asymmetries, whatever the exact nature of this $m=2$ asymmetry (central bar and or outer spiral arms).

\begin{figure}
\rotatebox{-0}{\includegraphics[width=8cm]{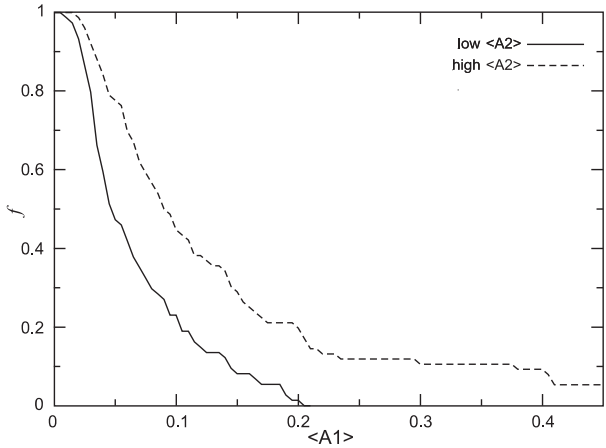}}
\caption{Cumulative function of $<A_1>$ for galaxies with values of $<A_2>$ higher than the median value of $<A_2>$ over our sample (dashed line), and $<A_2>$ smaller than this threshold (solid line). Galaxies with large $m=2$ asymmetries are clearly more lopsided than galaxies with small $m=2$ asymmetries. Here $<A_2>$ is the average value of $A_2$ between 1.5 and 2.5 disk scalelengths. The correlation between $A_1$ and $A_2$ remains if we take the average value of $A_2$ between 0.5 and 1.5 scalelengths (which would rather correspond to a central bar), or between 2.5 and 3.5 scalelengths (which would rather correspond to outer spiral arms).}
\label{A1A2}
\end{figure}

\subsection{Dependence of $A_1$ on the presence of companions \label{tidal}} 

To quantify the effect of tidal forces, we define
the tidal parameter as

\begin{equation}
T_{\mathrm{P}} = \log \left( \sum_i \frac{M_i}{M_0} \left( \frac{R_0}{D_i} \right) ^3 \right)
\end{equation}
where the sum is computed over the companions, $M_0$ is the mass of the galaxy considered, and
$R_0$ its scalelength. $M_i$ is the mass of each companion,
and $D_i$ its projected distance on the sky.
The ratio of masses is estimated from the ratio of blue
luminosities. The search of neighbors was done with the NED database, 
with a search radius of 2.5 degrees on the sky.
We take only the companions with measured radial velocities
within 500~km~s$^{-1}$.

The $<A_1>$ parameter values for the most lopsided galaxies are displayed versus the tidal parameter
in Fig.~9. No correlation is found. Strong lopsidedness with $<A_1>$ as large as 0.1 or larger is frequent in isolated galaxies (with $T_\mathrm{P}<-4$) without any close companion or any sign of a recent merger. NGC~1637 is a well-known example, but there are many other similar objects in our sample. The absence of correlation between strongly lopsided disks and the presence of companions agrees with what Wilcots \& Prescott (2004) found. Note that at this stage, it does not rule out interactions/mergers as being responsible for disk lopsidedness: the explanation could be that lopsidedness is very long-lived.

\begin{figure}
\rotatebox{-0}{\includegraphics[width=8cm]{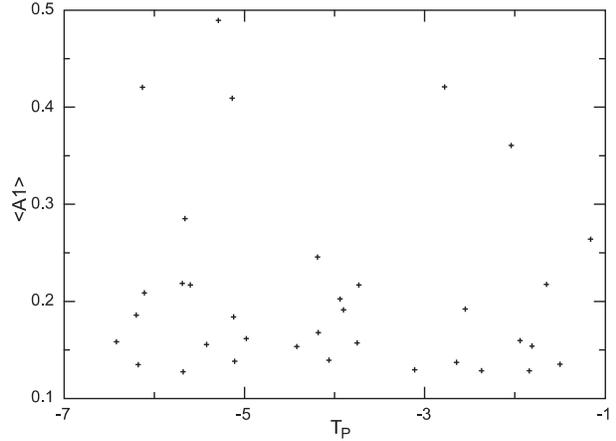}}
\caption{Correlation between the $<A_1>$ parameter and the tidal parameter $T_{\mathrm{P}}$ defined in text, for the 35 most lopsided galaxies with $i<70^\mathrm{o}$. No correlation is found. Strong lopsidedness in very isolated galaxies ($T_\mathrm{P}<-6$) is frequent, and is not the privilege of galaxies with companions. 
\label{A_1tidal}}
\end{figure}

\section{Origin of disk lopsidedness \label{simu}}

To understand the origin of lopsided disks, what must be explained is not only the frequency and strength of lopsidedness, but also the dependence on the Hubble type and $m=2$ asymmetries and the non-correlation of strong lopsidedness with the presence of interacting companions. 

\subsection{Scenarios for the origin of lopsidedness}
Minor $m=1$ perturbations can spontaneously arise in a disk. For instance, a two-arm spiral structure is never perfectly symmetrical, which results in a non-zero value of $<A_1>$. The numerical simulations detailed below, when run for isolated galaxies, show that the value of $<A_1>$ arising spontaneously in an isolated galaxy rarely exceeds 0.05 (in simulations of isolated galaxies, $<A_1>$ is smaller than this threshold 88\% of the time), and is never as large as 0.1 between 1.5 and 2.5 disk scale-lengths (the inner lopsidedness, at small radii, can be stronger, but is not studied in this paper). Thus, the lopsidedness observed in most spiral galaxies is the result of an external perturbation. Three scenarios to explain a galaxy's becoming significantly lopsided can then be proposed:
\begin{itemize}
\item distant interaction with a galaxy of comparable mass. The companion must have a high velocity and/or be distant enough for this encounter not to result in a merger; otherwise the merger remnant would be an elliptical galaxy, while we only have disk galaxies in our sample.
\item a minor merger with a small companion, with a mass ratio such as 10:1 or even more: such a companion is not massive enough to transform the system into an elliptical galaxy, but may induce important perturbations like lopsidedness. 
\item accretion of external gas. If the accretion is asymmetrical, this may result in a lopsided disk, even in the stellar component, when the gas is converted into stars.
\end{itemize}

We now study these three scenarios in numerical simulations, and compare the results of each scenario with the properties of the observed galaxy sample.

\subsection{Numerical techniques}

In our N-body code, galaxies are described by particles of  stars, gas, and dark-matter. We use $N=10^6$ particles. The gravitational interactions are computed with the three-dimensional FFT code of Bournaud \& Combes (2003) on $256^3$ grids, with a spatial gravitational softening of 300 pc. The dissipative nature of the ISM is modeled by the sticky-particles scheme of Bournaud \& Combes (2002), with elasticity parameters $\beta_r=\beta_t=0.8$. Star formation is described by a generalized Schmidt law (Schmidt 1959), and time-dependent stellar mass-loss is described by a model inspired by Jungwiert et al. (2001). More details about these codes can be found in Bournaud \& Combes (2002 and 2003).

Each spiral galaxy is modeled by a stellar and gaseous Toomre disk of scale-length 5~kpc and radius 15~kpc, of mass $10^{11}$~M$_{\sun}$, a central bulge, and a spherical dark halo. All the components are live (treated with self-gravitating particles) in these simulations. The bulge-to-disk and dark-to-visible mass ratios that we have used, as a function of the Hubble type (Sa--Sd), are given in Table~\ref{params}. The run parameters for each simulation are given later.

\begin{table}
\centering
\begin{tabular}{cccc}
\hline
\hline
Type & bulge-to-disk & halo-to-disk \\ 
     &   mass ratio  &  mass ratio  \\ 
\hline
 Sa  &     0.5       &     0.4      \\ 
 Sb  &     0.3       &     0.5      \\ 
 Sc  &     0.2       &     0.6      \\ 
 Sd  &     0.1       &     0.8      \\ 
\hline
\end{tabular}
\caption{Bulge, disk, and halo parameters used in numerical simulations. The 
halo mass that is accounted for is the mass of dark matter inside the disk 
radius.}
\label{params}
\end{table}

In these simulations we measure $<A_1>$, $<A_2>$, $Q_1$, and $Q_2$, at each time step, exactly as in observed systems, so that a robust comparison can be made. 

\subsection{Distant galactic encounters}

We first study distant encounters (without mergers) between galaxies of comparable masses. The parameters of the simulations that we ran are shown in Table~\ref{param_dist}. We focused on situations favorable to the formation of large lopsidedness in the disk. This is mainly the case on retrograde orbits, which is naturally explained by the fact that the pattern speed of $m=1$ asymmetries, $\Omega-\kappa$, is negative in a galactic disk, in contrast to $m=2$ spiral arms or bars that are more easily triggered on direct orbits, for $\Omega-\kappa/2$ is positive (e.g., G\'erin et al. 1990). 

\begin{figure}[!ht]
\centering
\resizebox{8cm}{!}{\includegraphics{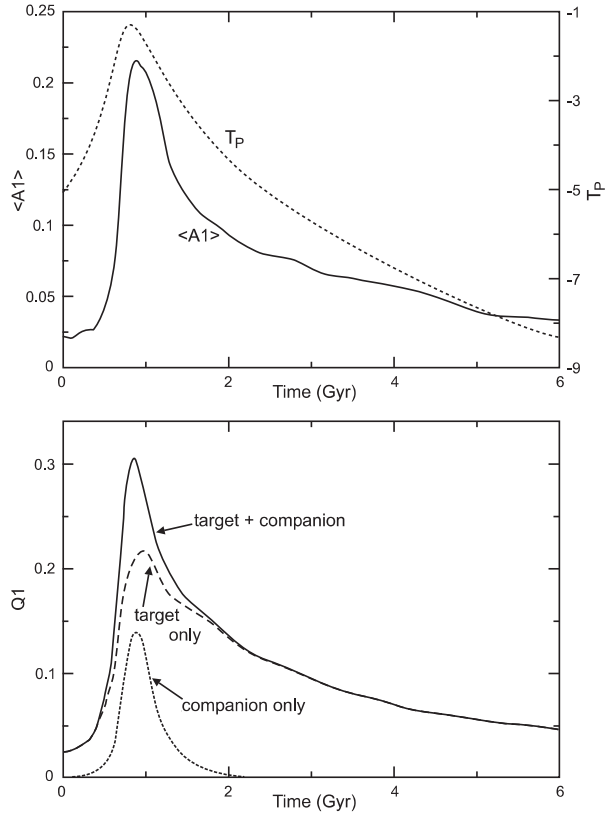}}
\caption{Simulation of distant interaction between two galaxies of comparable mass (Run~9). {\bf Top:} temporal evolution of $<A_1>$ (solid line) and the tidal parameter $T_{\mathrm{P}}$ (dashed, defined in Sect.~3). A strong lopsidedness is triggered during the interaction. It is observed for about 4 Gyrs, but is much lower once the system is isolated. -- {\bf Bottom:} $Q_1$ parameter in the disk plane for the total potential (solid line), the internal target galaxy potential only (dashed curve), and the external potential of the companion only (dotted curve). During the interaction, a strong lopsidedness results from the potential of the companion, but is not totally supported by the potential of the target galaxy itself. The maximal lopsidedness in $Q_1$ can be decomposed as an external contribution from the companion (short-lived) and an internal lopsidedness supported by the target galaxy potential (long-lived). This explains why there is a strong peak in $Q_1$ and $A_1$ during the interaction, and a long-lived but weaker lopsidedness after the interaction.}\label{sim_dist}
\end{figure}

\begin{table}
\centering
\begin{tabular}{lcccccc}
\hline
\hline
N$^\mathrm{o}$ & $M$ & $T$ & Orient. & $V$ (km s$^{-1}$) & $r$ (kpc) & $\theta$ \\
\hline
1  & 1:1 &Sb & D & 200 & 220 & 0   \\
2  & 1:1 &Sb & D & 300 & 130 & 0   \\
3  & 1:1 &Sb & D & 150 & 350 & 0   \\
4a & 1:1 &Sa & R & 250 & 130 & 0   \\
4b & 1:1 &Sb & R & 250 & 130 & 0   \\
4c & 1:1 &Sc & R & 250 & 130 & 0   \\
5a & 1:1 &Sa & R & 160 & 300 & 0   \\
5b & 1:1 &Sb & R & 160 & 300 & 0   \\
5c & 1:1 &Sc & R & 160 & 300 & 0   \\
6a & 1:1 &Sa & R & 200 & 220 & 0   \\
6b & 1:1 &Sb & R & 200 & 220 & 0   \\
6c & 1:1 &Sc & R & 200 & 220 & 0   \\
7  & 1:1 &Sb & R & 250 & 130 & 60  \\
8  & 1:2 &Sb & R & 250 & 130 & 0   \\
9  & 2:1 &Sb & R & 250 & 130 & 0   \\
10 & 1:3 &Sb & R & 250 & 130 & 0   \\
11 & 3:1 &Sb & R & 250 & 130 & 0   \\
12 & 1:2 &Sb & R & 160 & 300 & 0   \\
13 & 2:1 &Sb & R & 160 & 300 & 0   \\
14 & 1:2 &Sb & R & 450 & 450 & 0   \\
15 & 2:1 &Sb & R & 450 & 450 & 0   \\
\hline
\end{tabular}
\caption{Parameters for the simulations of distant interactions of galaxies. $M$ is the mass ratio (1:2 means that the companion is smaller than the target galaxy, 2:1 that it is more massive). $T$ is the Hubble Type (see Table~\ref{params}). The orientation of the orbit can be direct (D) or retrograde (R). We also give the relative velocity at an infinite distance $V$, the impact parameter $r$, and the inclination of the orbital plane with respect to the galactic plane $\theta$.}\label{param_dist}
\end{table}

\begin{figure}
\centering
\resizebox{8cm}{!}{\includegraphics{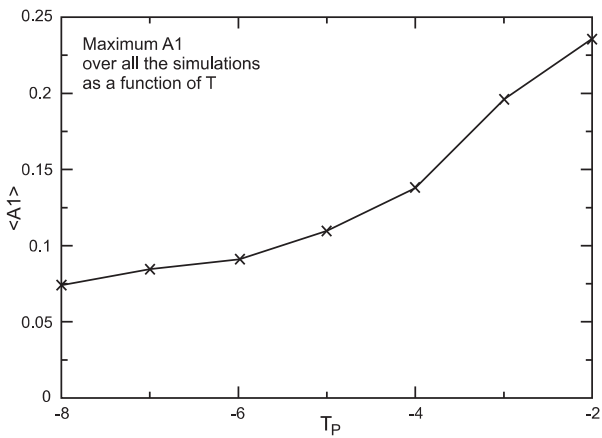}}
\caption{Largest value of $<A_1>$ as a function of the tidal parameter $T_{\mathrm{P}}$ for all the simulations of tidal galaxy interactions. All the results of the simulations lie below the solid line in this ($A_1$;$T_{\mathrm{P}}$) plane. This scenario does not explain the presence of strong lopsidedness ($<A_1>$ larger than 0.1) in isolated galaxies (tidal parameter $<-5$), while there are many observed cases.}\label{sim_distall}
\end{figure}

The temporal evolution of the $<A_1>$ parameter for a representative case (Run~9) is shown in Fig.~\ref{sim_dist}. During the interaction itself, a strong lopsidedness with values of $<A_1>$ higher than 0.15 is triggered. It is short-lived, lasting for about 500 Myrs. After that, a moderate but long-lived $m=1$ mode is observed during at least 4 Gyrs. All the simulations of distant galaxy interactions show the same behavior. This result is explained by analysing of the potential in terms of the $Q_1$ parameter, shown in Fig.~\ref{sim_dist}. The strong peak in $<A_1>$ and $Q_1$ observed during the interaction is supported for one part by the internal potential of the target galaxy, and for the other part by the external potential of the companion that pulls-up the disk. When the companion has left, only the part supported by the target galaxy potential remains; this mode is long-lived, but weaker than the peak observed in $Q_1$ and in $A_1$, during the interaction.

As a consequence of this temporal evolution, a strong lopsidedness is only observed when the companion is still close to the galaxy. This is the case in all the simulations, as shown in Fig.~\ref{sim_distall} with the largest value of $<A_1>$ over our whole simulation sample as a function of the tidal parameter. Even over the whole parameter space that we have covered, we never find a lopsidedness that is long-lived enough and strong enough to result in $<A_1>$ values higher than 0.1 in galaxies that appear very isolated. Strong lopsidedness, with $<A_1>$ of 0.15 or more, are found only when the tidal parameter $T_\mathrm{P}$ is larger than $-4$ in this scenario: depending on the mass ratio, this typically implies that the companion is within a few hundreds of kpc from the galaxy, and has not left the scene yet.

\subsection{Minor galaxy mergers}

\begin{figure}[!ht]
\centering
\resizebox{8cm}{!}{\includegraphics{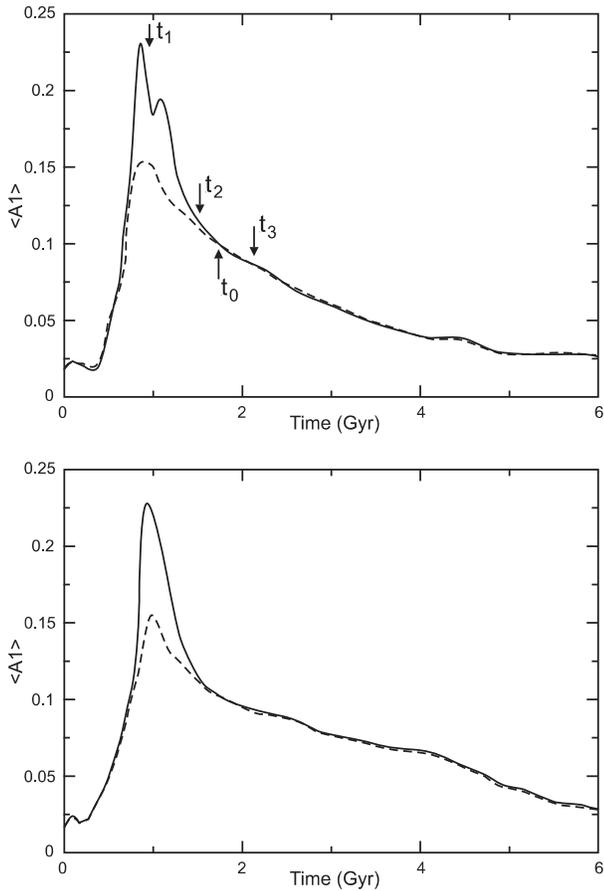}}
\caption{Temporal evolution of $<A_1>$ in two minor mergers (top: Run~4b, bottom: Run~6c). The solid line is when stars from the companion (or formed in gas from the companion) are included. The dashed line corresponds only to stars from the target galaxy (or formed in gas clouds present in the target before the merger). When the two curves have significantly different values, this means that the system is highly disturbed, for instance at $t_1$ (see the corresponding snapshot for Run~4b on Fig.~\ref{snap}). Tidal tails, streams, and other merger indicators, are still seen at $t_2$. The system can be classified as an ''isolated galaxy'' after the two curves have reached similar values, for instance at $t_3$ (see Fig.~\ref{snap}). We then consider that the maximal lopsidedness for an ''isolated'' galaxy is found when the two curves reach similar values, at $t_0$: corresponding values of $<A_1>$ are given for each simulation in Table~\ref{param_minor}.}
\label{sim_minor}
\end{figure}

\begin{table}
\centering
\begin{tabular}{lccccccc}
\hline
\hline
N$^\mathrm{o}$ & $M$ & $T$ & Orient. & $V$         & $r$ & $\theta$ & $<A_1>$ \\
               &     &   &         & km s$^{-1}$ & kpc &          & isolated  \\
\hline
1   & 10:1 & Sb & D & 100 & 50 & 0   &  0.08  \\
2a  & 10:1 & Sa & R & 100 & 50 & 0   &  0.11  \\
2b  & 10:1 & Sb & R & 100 & 50 & 0   &  0.12  \\
2c  & 10:1 & Sc & R & 100 & 50 & 0   &  0.11  \\
3   & 10:1 & Sb & D & 150 & 30 & 0   &  0.07  \\
4a  & 10:1 & Sa & R & 150 & 30 & 0   &  0.11  \\
4b  & 10:1 & Sb & R & 150 & 30 & 0   &  0.10  \\
4c  & 10:1 & Sc & R & 150 & 30 & 0   &  0.11  \\
5   & 10:1 & Sb & D &  60 & 40 & 0   &  0.09  \\
6a  & 10:1 & Sa & R &  60 & 40 & 0   &  0.11  \\
6b  & 10:1 & Sb & R &  60 & 40 & 0   &  0.13  \\
6c  & 10:1 & Sc & R &  60 & 40 & 0   &  0.12  \\
7   & 20:1 & Sb & D & 100 & 50 & 0   &  0.07  \\
8a  & 20:1 & Sa & R & 100 & 50 & 0   &  0.10  \\
8b  & 20:1 & Sb & R & 100 & 50 & 0   &  0.10  \\
8c  & 20:1 & Sc & R & 100 & 50 & 0   &  0.12  \\
9   & 20:1 & Sb & D & 150 & 30 & 0   &  0.08  \\
10  & 20:1 & Sb & R & 150 & 30 & 0   &  0.11  \\
11  & 10:1 & Sb & R &  60 & 40 & 60  &  0.09  \\
12  & 10:1 & Sb & R & 100 & 50 & 60  &  0.10  \\
13  & 20:1 & Sb & R & 200 & 30 & 0   &  0.09  \\
14  & 10:1 & Sb & R & 200 & 50 & 0   &  0.11  \\
15  &  7:1 & Sb & R &  60 & 40 & 0   &  0.13  \\
16  &  7:1 & Sb & R & 100 & 50 & 0   &  0.13  \\
17  &  7:1 & Sb & R & 150 & 30 & 0   &  0.14  \\
\hline
\end{tabular}
\caption{Parameters for the simulations of minor mergers. $M$ is the mass ratio. $T$ is the Hubble Type, determining several bulge/halo parameters (see Table~\ref{params} for details). The orientation of the orbit can be direct (D) or retrograde (R). We also give the relative velocity at an infinite distance $V$, the impact parameter $r$, and the inclination of the orbital plane on the galactic plane $\theta$. The last column indicates the value of $<A_1>$ when the stars of the companion are azimuthally mixed in the galaxy (i.e. when the value of $<A_1>$ is unchanged when we include these stars or not, at the instant called $t_0$ on Fig.~\ref{sim_minor}). Before that, the lopsidedness can be stronger, but the system appears like an ongoing merger, not like an isolated galaxy, as explained in text and shown in Figs.~\ref{sim_minor} and \ref{snap}. The value indicated here corresponds to the instant at which the galaxy begins to appear ''isolated''.}\label{param_minor}
\end{table}

The second scenario that we study is the merger with a minor companion. Walker et al. (1996) have already found that this process can form significant disk lopsidedness, visible for at least 1 Gyr in their simulations . The corresponding run parameters are given in Table~\ref{param_minor}. Again, the most favorable cases are on retrograde orbits. A mass ratio of 10:1 can trigger significant lopsidedness. Mergers with smaller mass ratios, such as 7:1, are even more efficient for triggering $m=1$ modes, but they will also lead to major disturbances of the system (see Bournaud et al. 2004, 2005), making the galaxy become an extremely early-type spiral or even an S0-like object, so they are not good candidates for the formation of lopsidedness in spiral galaxies in general, and we do not simulate mass ratios larger than 7:1. At the opposite, large mass ratios like 20:1 trigger weaker lopsidedness with typical values of $<A_1>$ smaller than 0.1. We have not studied mass ratios larger than 20:1 that cannot be the clue to the frequent lopsidedness with $<A_1> > 0.1$ that is observed.

The temporal evolution of $<A_1>$ in a minor merger (see Fig.~\ref{sim_minor} for runs 4b and 6c) is fairly similar to what occurs in distant galactic encounters without mergers. A very strong lopsidedness is triggered during the merger with values of $<A_1>$ that can be larger than 0.2, but this lasts typically no more than 500 Myrs. Later on, a long-lived $m=1$ mode is still present, but is weaker with values of $<A_1>$ not much larger than 0.1. The strong peak in $<A_1>$ observed during the merger is mainly related to the very asymmetric distribution of the stars from the companion; we find a lower value of $<A_1>$ when we only consider the target galaxy material (see Fig.~\ref{sim_minor}). After 2--3 rotations, the companion stars are mixed in the system and follow the same azimuthal distribution as the other stars -- even if their radial distribution is different. The value of $<A_1>$ is then smaller, but present in all the stars, regardless of whether stars from the companion are included in the analysis or not.

\begin{figure*}
\centering
\resizebox{16cm}{!}{\includegraphics{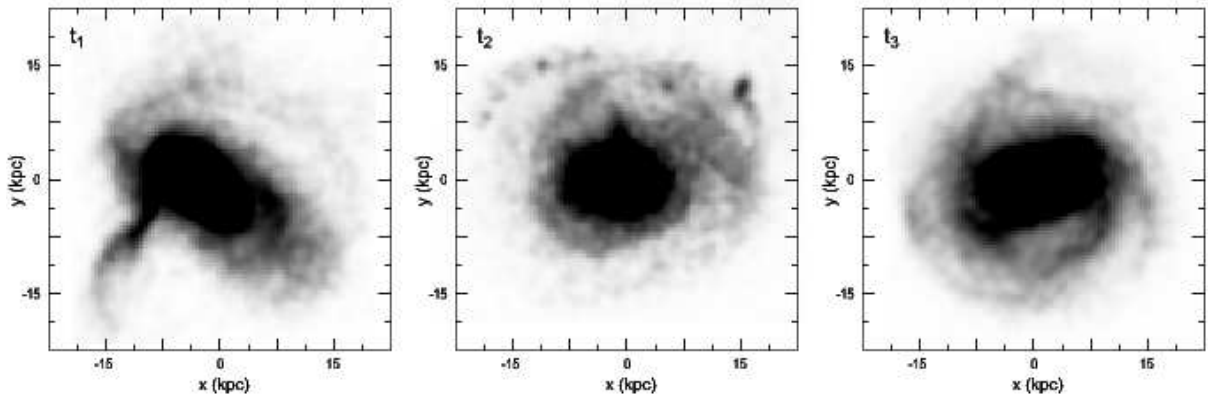}}
\caption{Snapshots of the minor merger scenario, Run~4b (see the temporal evolution of $<A_1>$ and the definition of $t_1$, $t_2$, $t_3$, in Fig.~\ref{sim_minor}). At $t_1$, there is a strong lopsidedness, but the ongoing merger is clearly visible. Even at $t_2$, when the stars from the companion are more ''mixed'' in the system, and no longer dominate the value of $<A_1>$, there are still obvious signs of the recent merger (like arcs and tidal debris in the outer parts, that are even more visible in the gas component). Only at $t_3$, the system is an ''isolated'' galaxy without sign of a past merger, but the amplitude of the lopsidedness is much lower, as shown by Fig.~\ref{sim_minor}.}
\label{snap}
\end{figure*}

Can minor mergers create a strong lopsidedness when the galaxy appears ''isolated'', once the most obvious merger-related features have disappeared? A strong lopsidedness does arise only during the merger, but the galaxy is then largely disturbed and would be classified as an ''on-going merger'', rather than as an ''isolated galaxy''. The question is then to know whether a strong lopsidedness can still be present when the merger is relaxed and the galaxy looks ''isolated''. As explained, the highest values of $<A_1>$ are observed when the azimuthal distribution of the stars of the companion is very different from the distribution of the other stars, at the instant called $t_1$ on Fig.~\ref{sim_minor}. This big difference in the two stellar components is enough to attest that a merger is occurring, and the visual aspect of the system clearly shows that it is an ongoing merger, not an isolated object (see Fig.~\ref{snap}). Even at time $t_2$, when including or not the stars from the companion does not largely change the value of $<A_1>$, there are still indicators that a recent merger has occurred, such as streams or tidal debris (see Fig.~\ref{snap}). It's only at time $t_3$, a few dynamical times later, that the system looks like an ''isolated'' galaxy -- but the value of $<A_1>$ is then twice lower than at $t_1$. Thus, the galaxy can look ''isolated'' only after we find similar values of $<A_1>$ with and without including the stars from the companion. Before this time, there is evidence that it is an ''ongoing merger''. Then, in Table~\ref{param_minor}, we included in the value of $<A_1>$ at the instant when the stars from the companion are azimuthally mixed in the system (called $t_0$ in Fig.~\ref{sim_minor}). The result is that, while the value of $<A_1>$ can be larger than 0.20 during the merger itself, the lopsidedness is much more modest once the merger remnant looks relaxed and can be classified as an ''isolated'' galaxy: at this moment, the value of $<A_1>$ is typically $\sim 0.10$ and is always smaller than 0.15, even for the not-so-minor 7:1 mergers  -- mergers with companions more massive than 7:1 are not considered, for the merger remnant is not a typical spiral-like system any more (see Bournaud et al. 2005).

\subsection{Difficulties with the galaxy interaction/merger scenarios}
Galaxy interactions, either distant encounters or minor mergers, can trigger strong lopsidedness. The galaxy can still be lopsided a few Gyrs after the event. However, several observational results cannot be explained by these scenarios.

\paragraph{Isolated galaxies that are strongly lopsided.}
As mentioned in Sect.~3, no correlation has been found between the strongest lopsidedness and the presence of interacting companions. Strong lopsidedness with $<A_1>\geq0.1$ are frequently found in galaxies that are very isolated and do not show any sign of a recent merger. 

This does not directly rule out galaxy interactions as responsible for lopsided disks, because a strong lopsidedness may {\it a priori} still be observed long after the interaction/merger, when no sign of it is observable any more. However, our simulations show that in the case of distant interactions with massive companions, the strongest lopsidedness can only be observed during the interaction, when the companion is still rather close, so that the strong lopsidedness with $<A_1> \geq 0.1$ in very isolated galaxies (tidal parameter of $-8$ to $-6$) that are frequently observed cannot be explained; on the contrary, with this scenario, a correlation of $<A_1>$ with the tidal parameter is expected, at least for the most lopsided galaxies. Minor mergers do not provide a better explanation, since the lopsidedness is very strong only during the merger itself, when the stellar population of the companion is not yet mixed in the disk, with other visible indicators that a merger has occurred recently (tidal tails, streams from the companions, etc.). Strong lopsidedness in galaxies that do not show any evidence of a recent merger and no massive interacting companion (like NGC~1637, M~101, and many others) cannot be accounted for. 

\begin{figure}
\centering
\resizebox{8cm}{!}{\includegraphics{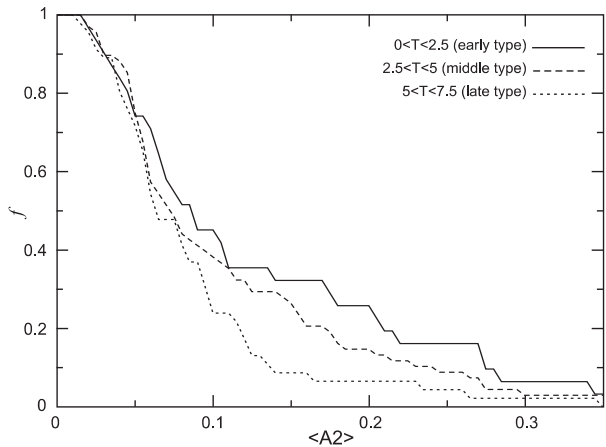}}
\caption{Cumulative function of $<A_2>$ for late-type, intermediate-type, and early-type spiral galaxies (a similar figure for $<A_1>$ has been shown in Fig.~\ref{hubbletype3}). Here $<A_2>$ is the average value of $A_2$ between 1.5 and 2.5 disk scalelengths, but the anti-correlation of $<A_2>$ with $T$ remains if we consider the average values between 0.5 and 1.5, or 2.5 and 3.5 disk scalelengths. Our sample obeys the general rule that $m=2$ bars/arms are stronger in early-type spiral galaxies. This is the opposite of what we found for lopsidedness, which is stronger in late-type galaxies.}\label{A2type}
\end{figure}

\begin{figure}
\centering
\resizebox{8cm}{!}{\includegraphics{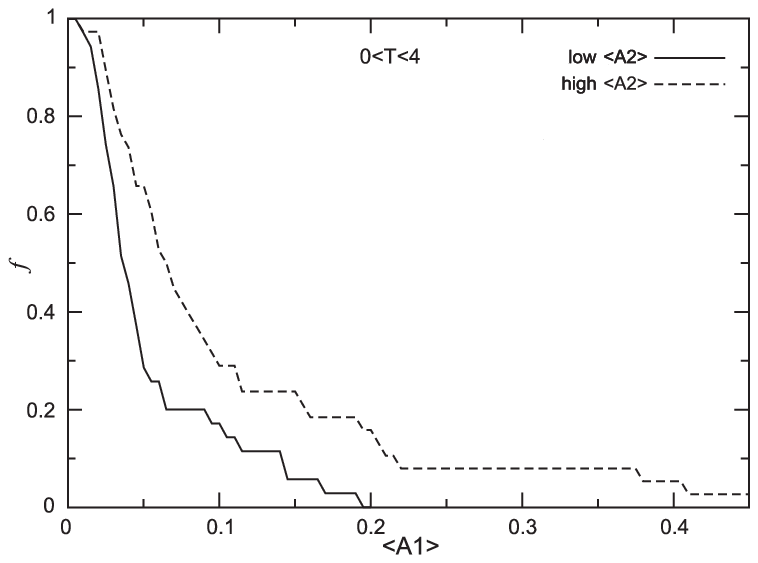}}\\
\resizebox{8cm}{!}{\includegraphics{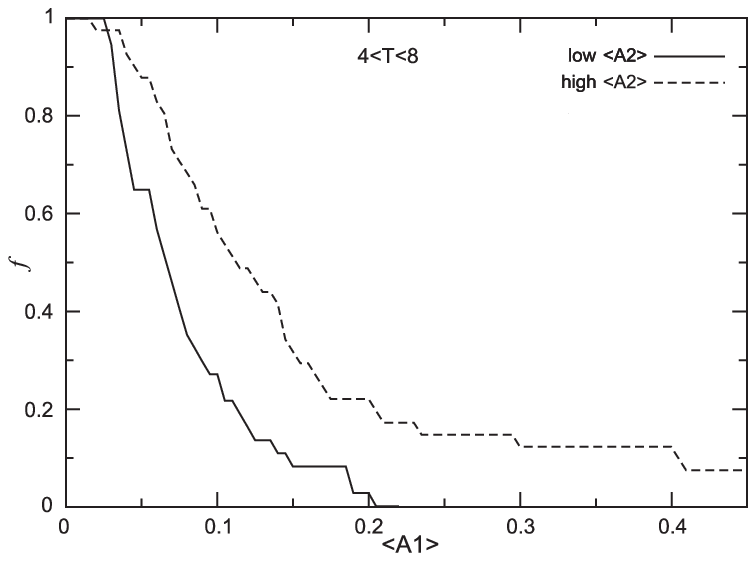}}
\caption{Same as Fig.~\ref{A1A2} for early-type spiral galaxies only (left, $0<T<4$) and late-type galaxies only (right, $4<T<8$). As discussed in the text, this shows that the correlation between $m=1$ and $m=2$ asymmetries found for the whole sample in Sect.~3 does not just mean that they are stronger in the same types of galaxies, but really means that they are preferentially formed simultaneously. Indeed, this correlation is still found when we limit the sample to early- or late-type galaxies.}\label{A1A2type}
\end{figure}

\paragraph{Correlation with $m=2$ asymmetries.}
We have shown in Sect.~3 that lopsidedness is correlated with the strength of $m=2$ asymmetries (spiral arms and/or bars). This result could have no relation with the physical origin of lopsidedness, for it could simply be that both lopsidedness and $m=2$ asymmetries are spontaneously stronger in the same type of galaxies. However there are at least two pieces of evidence that it is not the case: ({\it i}) lopsidedness is stronger in late-type galaxies (Sect.~3 and Fig.~7), while $m=2$ spiral arms/bars are stronger in early-type spiral galaxies (e.g. Elmegreen \& Elmegreen 1985, and also our own sample, see Fig.~\ref{A2type}), and ({\it ii}) when we take only early-type or late-type spirals, the dependence between $A_1$ and $A_2$ remains as strong as for the whole sample (see Fig.~\ref{A1A2type}). The correlation between $A_1$ and $A_2$ is thus related to the formation of lopsidedness and $m=2$ asymmetries. Yet, galaxy interactions/mergers do not explain it because: ({\it i}) the strongest lopsidedness are triggered on retrograde orbits, while ({\it ii}) $m=2$ asymmetries are triggered on prograde orbits; on retrograde orbits they are not triggered but can instead be weakened (e.g., G\'erin et al. 1990, Noguchi 1996). One would then expect weaker $m=2$ asymmetries in the most lopsided galaxies, which contradicts of the observational reality. The dependence observed between $A_1$ and $A_2$ cannot be explained if one assumes that galaxy interactions are responsible most of the observed lopsidedness.

\paragraph{Correlation with Hubble type.}
We have shown that there is a dependence of $A_1$ on the galaxy type: late-type spiral galaxies are much more lopsided than early-type ones. The same result has been found in gaseous disks by Matthews et al. (1998). This could mean that it is easier to trigger a strong lopsidedness in a late-type galaxy. However, simulations of interactions and mergers do not show significantly larger lopsidedness in late-types (there might be a small dependence on the Hubble type in simulations, but not as large as the observed one). Then, lopsidedness does not seem to be intrinsically stronger in late-type galaxies. The dependence on the observed Hubble Type must be a result of the formation mechanism, but it cannot be explained by galaxy interaction scenarios.

Lopsidedness is not only {\it stronger} in late-type galaxies, where there are more values of $<A_1> \; \sim 0.15 - 0.20$ than in early-type (see Fig.~7), they are also {\it more frequent} in late-type galaxies: if we choose the threshold for a galaxy to be regarded as lopsided at $<A_1>=0.05$, as suggested by our simulations, then 96\% of late-type galaxies are lopsided, while only 62\% of early-type spiral disks are lopsided (and this difference still holds if we chose another threshold). While galaxy interactions could have {\it a priori} explained that lopsidedness is stronger in late types (even when the simulations finally show they do not), they cannot explain at all why it is more frequent. On the contrary, a galaxy interaction makes the system become earlier-type. For the minor mergers, that are much more likely than distant equal-mass interactions, we have shown that the merger must not be too minor to trigger strong lopsidedness: mass ratios such as 10:1 or 7:1 are efficient, but 20:1 mergers are already much less efficient. According to our recent study of galaxy mergers (Bournaud et al. 2005), a 10:1 merger is enough to make the system evolve significantly towards early-type (more massive bulge, thicker disk, less gas), especially on retrograde orbits; and a 7:1 merger can even convert a typical spiral galaxy into a S0-like object. This evolution towards early-type is significant enough to be observable. Indeed, the variations in the bulge masses, velocity dispersions, and disk thickness, that result from a 10:1 merger correspond to a decrease in the Hubble type indicator $T$ by a few units. So, minor mergers would at the same time trigger an $m=1$ asymmetry and make the system become earlier-type. Distant galaxy interactions also would trigger lopsidedness and make the system become significantly earlier-type at the same time. 

Thus, the galaxy interaction/merger scenarios would tend to create more lopsidedness in early-type galaxies, because even if the galaxy is not early-type before the interaction, it will become early-type during the interaction itself, especially for the strongest lopsidedness for which a strong interaction/merger is required. An anti-correlation between $<A_1>$ and the Hubble type $T$ would then be expected if all the disk lopsidedness were the results of galaxy interactions/mergers. But the observational result is that lopsidedness is stronger and more frequent in late-type galaxies, which the interaction/merger scenarios fail to explain. 

\paragraph{Conclusion}
Several properties of lopsided galaxies, derived from our observed sample in Sect.~3, cannot be explained if one assumes that disk lopsidedness is mostly the result of galaxy interactions or mergers. 

We do not deny that distant galaxy interactions and minor mergers are responsible for the observed lopsidedness in some galaxies. This is shown by our simulations as well as previous works; and since interactions/mergers do occur in the Universe, they have triggered some of the observed lopsidedness. But we find that all the lopsided disks cannot be the result of these processes, for otherwise many observational properties could not be explained at all. We cannot give the exact fraction of lopsidedness that is not the result of galactic interactions, but it must be large enough to account for the overall statistical properties of our sample, so it is certainly several tens of percent. 

\subsection{Gas accretion}

\paragraph{Simulations of gas accretion}
Since a large fraction of lopsided disks are not the result of galaxy interactions and mergers and not the results of internal processes (that rarely lead to $<A_1>$ values larger than 0.05, as mentioned in Sect.~4.1), another scenario for making a galaxy lopsided must be proposed. We can then imagine that lopsidedness is the result of gas accretion. Standard cosmological models predict that galaxies accrete large amounts of gas along cosmological filaments (e.g., Keres et al. 2004). The mass accreted over a few billion years represents a significant fraction of the disk mass itself. This accretion not only fuels the gaseous disk, but also the stellar one, through star formation. It has already been shown that gas accretion fueling both the gaseous and stellar disk can be responsible both for the maintenance of spiral arms during a Hubble time, while spiral arms in isolated models generally do not last more than 2--3 Gyrs, and for the reformation of bars if they have been dissolved either by a merger or by an internal mechanism (Bournaud \& Combes 2002, Block et al. 2002). 

\begin{figure}
\centering
\resizebox{8cm}{!}{\includegraphics{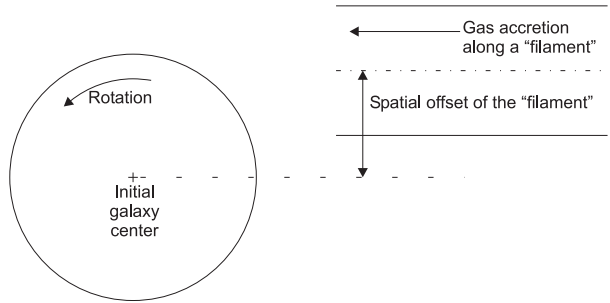}}
\caption{Configuration of accretion simulations with one filament. The spatial offset of the filament with respect to the disk center is given in Table~\ref{tab_acc} for each simulation. The size of the filament is 25 kpc.}\label{schem_accr}
\end{figure}

\begin{figure}
\centering
\resizebox{8cm}{!}{\includegraphics{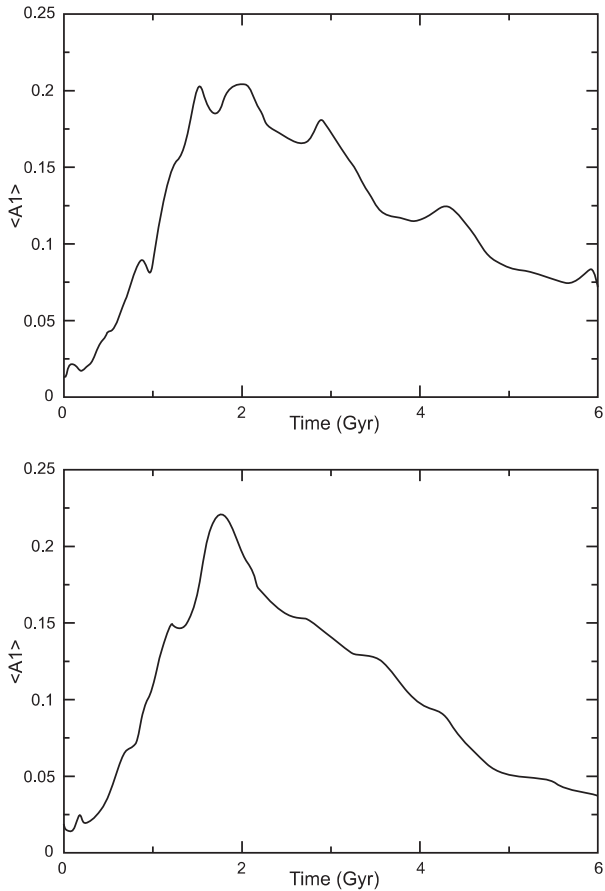}}
\caption{{\bf Top:} temporal evolution of $A_1$ in a simulation of accretion (Run 2). A strong lopsidedness is triggered, even in the stellar component (through star formation), in a galaxy that looks completely isolated. Then the value of $<A_1>$ decreases slowly while the disk re-orients itself, but remains larger than 0.1 during more than 3 Gyrs. -- {\bf Bottom:} same simulations with no accretion after $t=2$ Gyr. The lopsidedness is long-lived (about 3 Gyrs), but slightly shorter-lived than in the galaxy interaction/merger scenario, probably because perturbations to the halo are weaker in this case.}\label{sim_accr}
\end{figure}

This accretion can be asymmetrical, since distributed along a few filaments, and the accretion rate can be different for each filament. This can then trigger an $m=1$ asymmetry in the disk. In this paper we do not try to simulate ''realistic'' accretion conditions predicted in detail by any cosmological model, but simply to understand whether asymmetrical accretion of gas and subsequent star formation can create a significantly lopsided disk. To this aim, we first consider accretion along one single filament, off-centered with respect to the galaxy (see Fig.~\ref{schem_accr}), and will discuss more realistic cases with several filaments next. The velocity distribution of the accreted gas is assumed to be the same as in the outer disk. We ran simulations for accretion rates of 3, 6, and 9 M$_{\sun} \cdot$yr$^{-1}$. Typical values of $<A_1>$ in this model are given in Table~\ref{tab_acc}. The temporal evolution of $<A_1>$ is shown in Fig.~\ref{sim_accr} for two cases. 

\begin{table}
\centering
\begin{tabular}{lcccc}
\hline
\hline
N$^\mathrm{o}$ & T & $\dot{M}$ & offset & mean $A_1$\\
\hline
1   & Sb & 6  & 15 kpc & 0.12  \\
2   & Sb & 6  & 25 kpc & 0.16  \\
3   & Sb & 6  & 40 kpc & 0.15  \\
4   & Sb & 3  & 25 kpc & 0.11  \\
5a  & Sa & 9  & 25 kpc & 0.18  \\
5c  & Sc & 9  & 25 kpc & 0.21  \\
\hline
\end{tabular}
\caption{Parameters for simulations of accretion: initial Hubble type of the galaxy, accretion rate (in M$_{\sun}$~yr$^{-1}$), and spatial offset of the filament with respect to the disk center. The last column indicates the average value of $<A_1>$ between $t=2$ and $t=4$ Gyrs.}\label{tab_acc}
\end{table}

\begin{figure*}
\centering
\resizebox{12cm}{!}{\includegraphics{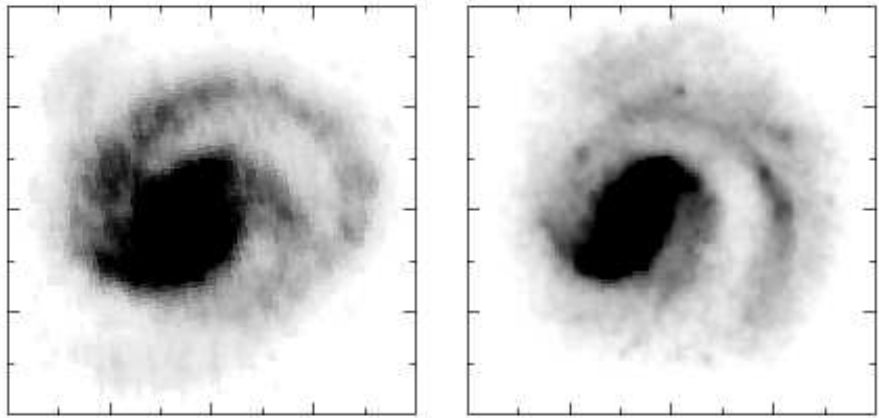}}
\caption{Snapshot of a simulation of asymmetrical gas accretion (left), giving an overall aspect rather similar to NGC~1637 (right: NIR map from the OSUBGS data, after deprojection). A strong lopsidedness is present in this very isolated galaxy. We have used two ''filaments'' to provide a more realistic case: the main filament shown in Fig.~\ref{schem_accr} is present, with an accretion rate of 4~M$_{\sun}$~yr$^{-1}$, and a second filament, symmetric of this first one with respect to the galaxy center, has an accretion rate of 2~M$_{\sun}$~yr$^{-1}$. These accretion rates are for a galactic mass of $10^{11}$~M$_{\sun}$, and correspond to doubling the mass of the galaxy in about one Hubble time.}\label{nac}
\end{figure*}

The main result of these simulations is thus that asymmetrical accretion of gas can create strong $m=1$ asymmetries, even in the stellar component after the formation of stars. The degree of lopsidedness decreases after a few Gyrs, even if accretion is not stopped, because the disk re-orients itself in regard to the filament. When accretion is very asymmetrical, as in our model with one filament, the values of $<A_1>$ can be as strong as in the extreme observed cases. This single-filament model is very arbitrary, and for a real galaxy it is more likely that there are several filaments with different accretion rates (e.g., Semelin \& Combes 2005). Then, the lopsidedness will generally be more moderate; however, it will remain true in such a realistic situation that asymmetrical gas accretion results in a lopsided disk. Indeed, we ran some simulations with 2 and 3 filaments and with different accretion rates for each filament: there is still a strong lopsidedness, that can reach values of $<A_1>$ larger than 0.10. We show for instance one case, that fits the morphology of NGC~1637, in Fig.~\ref{nac}. 

When accretion is stopped in the simulation, the lopsidedness has a typical lifetime of 3 Gyrs (see Fig.~\ref{sim_accr}). It is thus rather long-lived, but its lifetime seems a bit smaller than in the minor merger and distant interaction scenarios. This is maybe due to the smaller perturbations of the halo in the case of gas accretion, while the halo can help to make an $m=1$ mode long-lived (e.g., Jog 1997). However, a detailed understanding of the lifetime of lopsidedness in each scenario is beyond the scope of this paper. 

\paragraph{Lopsidedness and other properties}
Gas accretion can be the clue to explaining strong lopsidedness in galaxies that have no sign of recent interaction/merger. We have shown that disks do not remain strongly lopsided long enough for these situations to be explained by mergers or interactions that have occurred long ago. With the accretion of gas, we can explain such cases, as for instance NGC~1637, the morphology of which is well reproduced in one of our simulations with accretion along 2 filaments (see Fig.~\ref{nac}). If asymmetrical accretion of gas with high rates is frequent, strong lopsidedness in isolated galaxies can be frequent, too.

The relation between $A_1$ and $A_2$ is also a natural consequence of gas accretion, for this favors both lopsidedness and $m=2$ arms/bars at the same time (Bournaud \& Combes 2002). This scenario can thus explain why disks with strong $m=2$ asymmetries are more lopsided.

Gas accretion can also explain why there is a higher fraction of lopsidedness and stronger lopsidedness in late-type galaxies. First, late-type galaxies are generally less massive than early-type, so if they accrete gas at the same absolute rate, this can represent a large relative amount of mass and trigger a larger lopsidedness. Second, late-type galaxies contain more gas, which could be due to the fact that they have accreted gas more recently (but there are other possible explanations). Finally, gas accretion tends to make a galaxy evolve towards late-type (Bournaud \& Combes 2002), so the most lopsided galaxies in this scenario would naturally tend to be later-type, in contrast to what occurs with galaxy mergers that can trigger strong lopsidedness but make the galaxy become earlier-type at the same time. The correlation between $<A_1>$ and the Hubble type $T$ would thus be naturally explained in the frame of cosmological gas accretion.

Then, the observational results that were not explained correctly by the interaction and merger scenarios can be accounted for by gas accretion. Interactions and mergers can still produce lopsidedness in a large number of galaxies, but it is required that a large fraction of lopsided disks are the result of accretion to explain observational properties of galaxies. 

This is fully compatible with the high fraction of lopsided disks in HI (Sect.~\ref{HI}). Indeed, this observational result does not rule out that some lopsidedness can be caused by galaxy interactions or mergers, but at least it agrees with the accretion scenario, according to which gaseous disks should be at least as lopsided as stellar disks.

\section{Discussion and conclusion}\label{discu}

We measured the Fourier amplitudes of the $m=1$ component indicating lopsidedness in the stellar disk for a sample of 149 galaxies observed in the NIR, 149 of them suitable for the study of lopsidedness (inclined by less than 70 degrees). At least two thirds of these galaxies are found to be significantly lopsided, in the sense that the $m=1$ asymmetry is too large to simply result from internal mechanisms ($<A_1>$ larger than 0.05). The mean value of $<A_1>$ in our sample of galaxies is 0.11, and 34\% of the galaxies show values higher than this. We have shown that disks are more lopsided in late-type galaxies, s well as when the $m=2$ asymmetries (bars and spiral arms) are strong. However, the presence of a strongly lopsided disk is uncorrelated to the presence of interacting companions. Many isolated galaxies with no evidence of a recent interaction or merger are strongly lopsided.

Our numerical simulations seek to understand the physical origin of lopsidedness in spiral galaxies. We first studied distant tidal interactions between galaxies, and minor galaxy mergers (not major mergers, for they result in elliptical-like systems). The second scenario is much more likely. In both situations, strong lopsidedness can arise. Yet, both scenarios fail to explain the large frequency of strong lopsidedness in galaxies that are very isolated with no companion and no sign of a recent merger. They also contradict the observed correlations of the disk lopsidedness with the Hubble type, and with the presence of $m=2$ asymmetries. We then simulated an arbitrary model of accretion along one single filament, but also more realistic cases with several filaments, which can explain both the statistical properties of the sample and the morphology of some well-known cases.

Our study is based not only on the quantitative values of $<A_1>$, but also on qualitative relations between $<A_1>$ and other parameters. This is important, for the quantitative comparison of $<A_1>$ values alone could be questioned, as there might be some systematic differences between the density in simulations and the observed ''density'' computed from NIR images, especially if the mass-to-light ratio of stars varies with radius. However, this does not affect the dependence of $<A_1>$ on other physical quantities. Both results (values of $<A_1>$ and dependence on other parameters) led us to the same conclusion, which is therefore robust. 

As for the possibility that disk lopsidedness could simply result from internal mechanisms, suggested for instance by Masset \& Tagger (1997), we do not rule it out; but in our simulations, as well as in many other numerical works in the literature, we have never observed strong lopsidedness arising spontaneously in disks. Internal mechanisms are found to lead only to values of $<A_1>$ typically smaller than 0.05, so they are not the clue to the origin of the observed disk lopsidedness\footnote{Internal mechanisms are not found to lead to high values of $<A_1>$ between 1.5 and 2.5 disk scale-length, but they can trigger stronger lopsidedness at smaller radii. They may be responsible for strong inner lopsidedness, but cannot account for the large-scale lopsidedness studied in this paper}, since $<A_1>$ is larger than 0.05 in most spiral galaxies. Internal mechanisms could only explain the weakest lopsidedness in our sample.

 To explain the observations, we show that a large fraction of lopsided disks result from accretion of gas, with a typical accretion rate that corresponds to doubling the disk mass in about 10 Gyrs, in agreement with the predictions by Block et al. (2002). Observations of lopsidedness in HI disks, as traced by the lopsidedness of the global HI profiles, agree with this accretion scenario, but NIR observations of stellar disks enable a better quantitative study than HI velocity profiles without spatial resolution.

Large accretion rates of a few M$_{\sun}\cdot$yr$^{-1}$ along cosmological filaments may explain both the ubiquity of lopsidedness (this paper) and of spiral arms and bars (Bournaud \& Combes 2002 and Block et al. 2002). In both cases of $m=2$ and $m=1$ asymmetries star formation makes the effects of gas accretion visible in the stellar component. The results presented here for lopsidedness can even be considered as more robust than our previous study based on bars. First, deprojection affects bars more than lopsidedness. Galactic disks are never perfectly round, even in their outer parts, so there is an uncertainty on the disk inclination, and the associated error in the deprojection will bias the $m=2$ asymmetry, but not the $m=1$ one; so this observational concern is absent in the present study. Second, the present conclusions are supported not only by the quantitative values of the asymmetry, but also by correlations with other physical parameters, providing several independent proofs of gas accretion. Third, lopsidedness mainly concerns large-scale outer disks, while bars are coupled with inner gas inflows and nuclear bars. The relative resolution of simulations is higher in the outer disk, so results on disk lopsidedness are likely to be more robust than results concerning bars.  Then, we find a robust confirmation of the result inferred from the study of bars in Block et al. (2002): without denying the role of interactions and mergers in the evolution of galaxies, we here find a new evidence that large-scale gas accretion must play a major role in the morphological properties of galaxies. 

\appendix

\section{$m=1$ and $m=2$ asymmetries in the OSUBGS sample}\label{app}
This appendix gives the values of $<A_m>$ for $m=1$ and $2$, as defined in Sect.~\ref{a1def}, for each galaxy in our sample, and the inclination angle $i$ used to compute these values.

\begin{table}
\centering
\begin{tabular}{lccc}
\hline
\hline
Name & $i$ & $<A_1>$ & $<A_2>$ \\
\hline
IC~5325 & 47.09 & 0.035 & 0.026 \\
NGC~150 & 63.70 & 0.043 & 0.130 \\
NGC~157 & 55.01 & 0.111 & 0.079 \\
NGC~210 & 53.66 & 0.050 & 0.072 \\
NGC~278 &  3.24 & 0.036 & 0.022 \\
NGC~289 & 39.50 & 0.093 & 0.059 \\
NGC~428 & 51.14 & 0.145 & 0.037 \\
NGC~488 & 43.32 & 0.034 & 0.018 \\
NGC~578 & 53.72 & 0.100 & 0.058 \\
NGC~613 & 49.77 & 0.018 & 0.110 \\
NGC~625 & 90.00 & 0.075 & 0.030 \\
NGC~685 & 34.21 & 0.036 & 0.053 \\
NGC~779 & 73.75 & 0.029 & 0.070 \\
NGC~864 & 47.94 & 0.123 & 0.090 \\
NGC~908 & 67.90 & 0.096 & 0.038 \\
NGC~988 & 61.30 & 0.066 & 0.073 \\
NGC~1003 & 75.70 & 0.242 & 0.022 \\
NGC~1042 & 61.10 & 0.055 & 0.083 \\
NGC~1058 & 45.36 & 0.090 & 0.017 \\
NGC~1073 & 38.28 & 0.168 & 0.139 \\
NGC~1084 & 52.20 & 0.041 & 0.030 \\
NGC~1087 & 54.19 & 0.060 & 0.075 \\
NGC~1187 & 48.50 & 0.147 & 0.048 \\
NGC~1241 & 56.86 & 0.101 & 0.075 \\
NGC~1300 & 58.81 & 0.096 & 0.103 \\
NGC~1302 & 27.49 & 0.030 & 0.054 \\
NGC~1309 & 15.95 & 0.154 & 0.048 \\
NGC~1317 & 31.99 & 0.020 & 0.046 \\
NGC~1350 & 64.10 & 0.028 & 0.086 \\
NGC~1371 & 51.30 & 0.048 & 0.050 \\
NGC~1385 & 49.00 & 0.232 & 0.130 \\
NGC~1421 & 83.47 & 0.102 & 0.119 \\
NGC~1493 & 20.31 & 0.055 & 0.049 \\
NGC~1511 & 79.83 & 0.114 & 0.064 \\
NGC~1559 & 58.15 & 0.186 & 0.030 \\
NGC~1617 & 68.31 & 0.079 & 0.040 \\
NGC~1637 & 31.73 & 0.186 & 0.070 \\
NGC~1703 & 20.49 & 0.144 & 0.019 \\
NGC~1792 & 65.09 & 0.116 & 0.034 \\
NGC~1808 & 70.19 & 0.281 & 0.090 \\
NGC~1832 & 61.48 & 0.056 & 0.057 \\
NGC~1964 & 69.39 & 0.063 & 0.074 \\
NGC~2090 & 63.79 & 0.201 & 0.033 \\
NGC~2139 & 40.38 & 0.086 & 0.079 \\
NGC~2196 & 40.98 & 0.066 & 0.036 \\
NGC~2280 & 64.24 & 0.154 & 0.076 \\
NGC~2442 & 49.88 & 0.196 & 0.127 \\
NGC~2566 & 60.19 & 0.054 & 0.102 \\
NGC~2775 & 40.42 & 0.195 & 0.019 \\
NGC~2964 & 53.53 & 0.093 & 0.094 \\
NGC~3059 & 21.40 & 0.409 & 0.073 \\
\hline
\end{tabular}
\end{table}

\begin{table}
\centering
\begin{tabular}{lccc}
\hline
\hline
Name & $i$ & $<A_1>$ & $<A_2>$ \\
\hline
NGC~3166 & 66.57 & 0.204 & 0.162 \\
NGC~3169 & 62.82 & 0.043 & 0.035 \\
NGC~3223 & 51.37 & 0.110 & 0.037 \\
NGC~3227 & 58.84 & 0.209 & 0.076 \\
NGC~3261 & 43.05 & 0.113 & 0.079 \\
NGC~3275 & 43.06 & 0.055 & 0.057 \\
NGC~3319 & 59.12 & 0.596 & 0.122 \\
NGC~3338 & 54.53 & 0.076 & 0.023 \\
NGC~3423 & 35.66 & 0.146 & 0.026 \\
NGC~3504 & 26.17 & 0.036 & 0.079 \\
NGC~3507 & 30.22 & 0.047 & 0.061 \\
NGC~3511 & 81.96 & 0.055 & 0.035 \\
NGC~3513 & 46.45 & 0.066 & 0.084 \\
NGC~3583 & 55.18 & 0.020 & 0.100 \\
NGC~3593 & 76.26 & 0.018 & 0.012 \\
NGC~3596 &  8.79 & 0.161 & 0.059 \\
NGC~3646 & 60.57 & 0.400 & 0.037 \\
NGC~3675 & 58.37 & 0.020 & 0.028 \\
NGC~3681 & 33.17 & 0.030 & 0.027 \\
NGC~3684 & 50.58 & 0.533 & 0.032 \\
NGC~3686 & 40.91 & 0.078 & 0.055 \\
NGC~3726 & 49.07 & 0.080 & 0.058 \\
NGC~3810 & 47.78 & 0.044 & 0.036 \\
NGC~3877 & 82.93 & 0.059 & 0.093 \\
NGC~3885 & 78.93 & 0.022 & 0.093 \\
NGC~3887 & 39.77 & 0.163 & 0.064 \\
NGC~3893 & 59.14 & 0.139 & 0.056 \\
NGC~3938 & 12.98 & 0.067 & 0.020 \\
NGC~3949 & 56.37 & 0.072 & 0.022 \\
NGC~4027 & 48.96 & 0.201 & 0.113 \\
NGC~4030 & 42.19 & 0.030 & 0.034 \\
NGC~4051 & 28.82 & 0.107 & 0.058 \\
NGC~4062 & 68.40 & 0.025 & 0.030 \\
NGC~4100 & 78.45 & 0.039 & 0.052 \\
NGC~4111 & 90.00 & 0.031 & 0.139 \\
NGC~4123 & 47.68 & 0.047 & 0.115 \\
NGC~4136 & 21.98 & 0.077 & 0.040 \\
NGC~4138 & 61.62 & 0.018 & 0.018 \\
NGC~4145 & 55.15 & 0.038 & 0.027 \\
NGC~4151 & 39.99 & 0.082 & 0.111 \\
NGC~4178 & 90.00 & 0.087 & 0.106 \\
NGC~4212 & 53.60 & 0.081 & 0.030 \\
NGC~4242 & 52.28 & 0.032 & 0.039 \\
NGC~4254 & 29.07 & 0.122 & 0.040 \\
NGC~4293 & 66.52 & 0.030 & 0.087 \\
NGC~4303 & 19.29 & 0.127 & 0.073 \\
NGC~4314 & 19.13 & 0.066 & 0.164 \\
NGC~4388 & 90.00 & 0.058 & 0.061 \\
NGC~4394 & 19.47 & 0.044 & 0.123 \\
NGC~4414 & 56.49 & 0.035 & 0.025 \\
NGC~4448 & 60.05 & 0.033 & 0.145 \\
NGC~4450 & 42.65 & 0.023 & 0.067 \\
NGC~4457 & 33.95 & 0.219 & 0.030 \\
NGC~4487 & 55.13 & 0.056 & 0.025 \\
NGC~4490 & 66.21 & 0.172 & 0.043 \\
NGC~4504 & 52.47 & 0.042 & 0.048 \\
NGC~4527 & 75.02 & 0.034 & 0.073 \\
NGC~4548 & 35.02 & 0.062 & 0.128 \\
NGC~4571 & 25.93 & 0.027 & 0.011 \\
NGC~4579 & 38.82 & 0.037 & 0.085 \\
NGC~4580 & 48.43 & 0.022 & 0.027 \\
\hline
\end{tabular}
\end{table}

\begin{table}
\centering
\begin{tabular}{lccc}
\hline
\hline
Name & $i$ & $<A_1>$ & $<A_2>$ \\
\hline
NGC~4593 & 54.91 & 0.027 & 0.062 \\
NGC~4618 & 44.54 & 0.299 & 0.083 \\
NGC~4643 & 29.92 & 0.057 & 0.157 \\
NGC~4651 & 49.97 & 0.031 & 0.018 \\
NGC~4654 & 57.69 & 0.143 & 0.029 \\
NGC~4665 &  0.00 & 0.009 & 0.127 \\
NGC~4666 & 74.57 & 0.069 & 0.019 \\
NGC~4689 & 39.39 & 0.030 & 0.014 \\
NGC~4698 & 58.75 & 0.028 & 0.067 \\
NGC~4772 & 63.29 & 0.026 & 0.120 \\
NGC~4775 & 27.56 & 0.089 & 0.029 \\
NGC~4781 & 67.95 & 0.105 & 0.074 \\
NGC~4818 & 81.59 & 0.072 & 0.128 \\
NGC~4856 & 90.00 & 0.051 & 0.200 \\
NGC~4902 & 26.47 & 0.094 & 0.089 \\
NGC~4930 & 45.75 & 0.818 & 0.125 \\
NGC~4939 & 67.43 & 0.044 & 0.041 \\
NGC~4941 & 53.51 & 0.021 & 0.016 \\
NGC~5005 & 67.11 & 0.026 & 0.081 \\
NGC~5054 & 54.25 & 0.057 & 0.030 \\
NGC~5078 & 90.00 & 0.049 & 0.060 \\
NGC~5085 & 35.55 & 0.074 & 0.043 \\
NGC~5101 & 40.28 & 0.156 & 0.146 \\
NGC~5121 & 40.35 & 0.027 & 0.025 \\
NGC~5161 & 70.70 & 0.064 & 0.125 \\
NGC~5247 & 42.86 & 0.096 & 0.077 \\
NGC~5248 & 50.43 & 0.072 & 0.048 \\
NGC~5334 & 42.47 & 0.037 & 0.029 \\
NGC~5371 & 47.78 & 0.490 & 0.100 \\
NGC~5427 & 41.98 & 0.137 & 0.059 \\
NGC~5448 & 64.60 & 0.060 & 0.082 \\
NGC~5483 & 23.82 & 0.208 & 0.047 \\
NGC~5676 & 65.62 & 0.104 & 0.023 \\
NGC~5701 & 29.31 & 0.408 & 0.103 \\
NGC~5713 & 39.74 & 0.110 & 0.117 \\
NGC~5838 & 90.00 & 0.024 & 0.139 \\
NGC~5850 & 41.42 & 0.034 & 0.097 \\
NGC~5921 & 44.52 & 0.042 & 0.117 \\
NGC~5962 & 49.29 & 0.056 & 0.052 \\
NGC~6753 & 30.98 & 0.031 & 0.017 \\
NGC~6782 & 41.99 & 0.111 & 0.114 \\
NGC~6902 & 43.42 & 0.041 & 0.039 \\
NGC~6907 & 44.98 & 0.169 & 0.157 \\
NGC~7083 & 57.12 & 0.043 & 0.037 \\
NGC~7184 & 84.57 & 0.030 & 0.151 \\
NGC~7205 & 61.75 & 0.063 & 0.026 \\
NGC~7213 & 28.66 & 0.014 & 0.004 \\
NGC~7217 & 33.86 & 0.015 & 0.017 \\
NGC~7412 & 40.61 & 0.074 & 0.070 \\
NGC~7418 & 40.82 & 0.034 & 0.055 \\
NGC~7479 & 41.81 & 0.088 & 0.167 \\
NGC~7552 & 51.37 & 0.085 & 0.189 \\
NGC~7582 & 69.88 & 0.021 & 0.084 \\
NGC~7606 & 60.08 & 0.049 & 0.023 \\
NGC~7713 & 65.41 & 0.141 & 0.024 \\
NGC~7723 & 49.53 & 0.143 & 0.062 \\
NGC~7727 & 40.02 & 0.375 & 0.042 \\
NGC~7741 & 48.17 & 0.067 & 0.146 \\
NGC~7814 & 67.29 & 0.017 & 0.040 \\
\hline
\end{tabular}
\caption{Inclination and values of $<A_1>$ and $<A_2>$ for galaxies in our sample.}\label{TabA}
\end{table}

\begin{acknowledgements}
We are grateful to the referee, Wim van Driel, whose constructive comments helped to improve the presentation of our results. The computations in this work have been realized on the Fujitsu NEC-SX5 of the CNRS computing center, at IDRIS. We are happy to acknowledge the support of the Indo-French grant IFCPAR/2704-1.
\end{acknowledgements}

\end{document}